\documentstyle[fleqn,prl,aps,preprint]{revtex}

\begin{document}
\draft
\widetext

\title{A new class of exactly solvable interacting fermion models in one
dimension} \author{H.J. Schulz} \address{Institute for Theoretical
Physics,
University of California, Santa Barbara, CA 93106 \\ Laboratoire de
Physique
des Solides, Universit\'e Paris--Sud, 91405 Orsay, France}
\author{B.S. Shastry} \address{Indian Institute of Science, Bangalore
560012, India} \maketitle

\begin{abstract}
We investigate a model containing two species of one--dimensional
fermions
interacting via a gauge field determined by the positions of all
particles
of the opposite species. The model can be solved exactly via a simple
unitary transformation. Nevertheless, correlation functions exhibit
nontrivial interaction--dependent exponents. A similar model defined on
a
lattice is introduced and solved. Various generalizations, e.g. to the
case
of internal symmetries of the fermions, are discussed. The present
treatment
also clarifies certain aspects of Luttinger's original solution of the
``Luttinger model''.
\end{abstract}

\narrowtext Exactly solvable models
\cite{bethe_xxx,luttinger_model,mattis_lieb_bos,lieb_hubbard_exact} have
played an important role in the current understanding of
one--dimensional
interacting many--particle systems. These together with the idea of
dominant
low energy bosonic excitations of Fermi systems \cite{tomonaga} gave
rise to
the emergence of the ``Luttinger liquid'' \cite{haldane_bosonisation} as
a
unifying concept. Nevertheless, the technicalities of these exact
solutions
(bosonization, Bethe ansatz) are often rather complex. In the present
paper
we wish to introduce a class of interacting models which can be
diagonalized
by a simple (pseudo--)unitary transformation, yet exhibit nontrivial
Luttinger--liquid behavior. The models can be defined both in the
continuum
and on a lattice, and can have rather arbitrary single--particle
bandstructure. Only the interactions are constrained to be of a
particular
``gauge form''. The long--distance asymptotics of correlation functions
can
then be determined either by direct calculation or by a mapping on an
effective Luttinger liquid description. Our investigation was inspired
by
Luttinger's original treatment of the ``Luttinger model'', and we will
comment on this connection further below.

We start by considering the simplest model in our class, a
one--dimensional
fermion model with two species of particles, designated by a pseudospin
index $\sigma = \pm$, having coordinates $x_{\sigma i}$ and momenta
$p_{\sigma i} = -i \partial_{x_{\sigma i}}$. The Hamiltonian of our
model
then is
\begin{equation} 
H = \frac12 \sum_{\sigma i} \Pi_{\sigma i}^2 
\label{eq:h}
\end{equation}
where we have introduced a ``covariant momentum'' $\Pi_{\sigma i} =
p_{\sigma i} + \sigma A_\sigma(x_{\sigma i})$, i.e. in this model, {\em
particles interact via a gauge potential, given for a particle at $x$ by
$A_\sigma(x) = \sum_j V(x-x_{-\sigma j})$}. The potential $V$ is an even
function and vanishes at infinity. On a ring of length $L$, we will
assume a
potential $V$ that is periodic in $L$.  Clearly, the Hamiltonian is not
time--reversal invariant, but it is invariant under simultaneous time
reversal and charge ($\sigma$) conjugation.

The model can now be straightforwardly diagonalized by a (in general
pseudo--)unitary transformation: noting that
\begin{equation} 
e^{i S(\{x_{+i}\},\{x_{-j}\})} p_{+i} e^{-i S(\{x_{+i}\},\{x_{-j}\})}
= p_{+i} - \partial_{x_{+i}} S(\{x_{+i}\},\{x_{-j}\})
\label{eq:uni}
\end{equation}
one can chose $S$ so as to eliminate the interaction in eq.(\ref{eq:h})
by 
\begin{equation} 
S(\{x_{+i}\},\{x_{-j}\}) =  \sum_{i,j} E(x_{+i}-x_{-j}) \;\;,
\label{eq:s}
\end{equation}
where $E$ is the indefinite integral of the interaction potential:
\begin{equation} 
E(x) = \int^x_0 dx' \; V(x') \;\;.
\label{eq:e}
\end{equation}
The transformed Hamiltonian then takes the form
\begin{equation} 
\tilde{H} = e^{i S} H e^{-i S}
= \frac12  \sum_{\sigma i} p_{\sigma i}^2  \;\;.
\label{eq:ht}
\end{equation}
The eigenfunctions of $\tilde{H}$ clearly are Slater determinants of
plane
wave states $|\{k_{-,i}\},\{k_{+,j}\}\rangle$, characterized by the sets
of
wavenumbers $\{k_{-,i}\}$ and $\{k_{+,j}\}$ for the $-$ and $+$
particles
respectively. Consequently, the eigenfunctions and eigenvalues of the
original Hamiltonian are obtained straightforwardly:
\begin{equation} 
H e^{-i S} |\{k_{-,i}\},\{k_{+,j}\}\rangle =
\frac12 \sum_{\sigma i} k_{\sigma i}^2 
e^{-i S} |\{k_{-,i}\},\{k_{+,j}\}\rangle \;\;.
\label{eq:heig}
\end{equation}
At first sight it thus appears that the spectrum of the interacting
Hamiltonian is independent of the interaction. Conformal field theory,
or
equivalently Luttinger liquid theory, then would imply that the
asymptotic
form of correlation functions (which is directly determined by the
eigenvalue spectrum) is also interaction--independent. This conclusion
is
however incorrect: periodic boundary condition have to be treated
carefully. In fact, keeping all other coordinates fixed, one easily
finds $
S(x_{-,i}=L) - S(x_{-,i}=0) = - N_+ \delta$, where $L$ is the length
over
which periodic boundary conditions are applied, $N_+$ is the total
number of
$+$ particles, and the {\em phase shift} $\delta$ is given by
\begin{equation} 
\delta = \int_{0}^L dx \; V(x) 
\label{eq:d}
\end{equation}
An analogous result, with $N_+$ and $\delta$ replaced by $N_-$ and
$-\delta$, holds for the phase shift of the $+$ particles. Consequently,
the
quantization condition on the wavenumbers is given by
\begin{equation} 
L k_{\pm,i} \mp N_\mp \delta = 2 \pi n_{\pm,i} \;\;,
\label{eq:q}
\end{equation}
where the $n_{\pm,i}$ are integer quantum numbers analogous to those
used in
the noninteracting case. Clearly, particles of one given ``spin''
orientation give rise to an effective Aharonov--Bohm flux acting on the
other species, the value of the flux depending on the number of
particles
present. It should now also be clear why we refer to the transformation
eq.(\ref{eq:uni}) as {\em pseudo}unitary: unless $\delta$ is
``accidentally'' an integer multiple of $2\pi$, the plane wave states of
the
interacting and noninteracting problems obey different boundary
conditions
and therefore define different Hilbert spaces.

The ground state energy $E_0$ can be found in any sector with $N_{\pm}$
particles, as follows. In order to minimize $E_0$ we must choose
$n_{\pm,i}=
n_{\pm,i}^0 \pm [[\frac{\delta}{2 \pi} N_{\mp}]]_{int}$, where
$n_{\pm,i}^0$
are the quantum numbers in the absence of the interaction, and we denote
$x=
[[x]]_{int}+[[x]]_{rem}$ for any $x$ where $[[x]]_{int}$ is the closest
integer to $x$. Thus $-1/2 \leq [[x]]_{rem} \leq 1/2$ for any $x$.  The
change in energy due to the interaction, $\delta E= \frac{2 \pi^2}{L^2}
\{
N_+ [[\frac{N_- \delta}{ 2 \pi}]]_{rem}^2 + N_- [[\frac{N_+ \delta}{ 2
\pi}]]_{rem}^2 \} $, is not extensive, but on the scale expected from a
magnetic field applied to the ring.

An effective Luttinger liquid description in terms of a bosonic field
theory
for the low--energy properties can be obtained from the low--energy
excited
states.\cite{haldane_bosonisation} To be precise, we start from a ground
state with $N_{\pm 0} = 2 n_0 +1$ and assume that $N_{\pm 0}\delta$ is
an
integer multiple of $2\pi$. We now add $n_{\pm R}$ ($n_{\pm L}$)
particles
at the right (left) Fermi points of the $\pm$ particles. Introducing
particle number and current quantum numbers $N_\pm = n_{\pm R} + n_{\pm
L}$
and $J_\pm = n_{\pm R} - n_{\pm L}$ the second order variation of the
ground
state energy is
\begin{equation} 
E^{(2)} = \frac{1}{2L^2} (2n_0 +1) \left[
(\pi^2+\delta^2)(N_+^2+N_-^2) + \pi^2(J_+^2+J_-^2)
+2\pi\delta(J_+N_- - J_-N_+)\right] \;\;.
\label{eq:h2}
\end{equation}
Up to quantum fluctuations, $N_\pm$ and $J_\pm$ are related to bosonic
fields and their conjugate momentum density via $ N_\pm = -(L/\pi)
\partial_x \phi_\pm$ and $J_\pm = L \Pi_\pm$. The effective Hamiltonian
including the low-energy quantum fluctuations then takes the form
\begin{eqnarray}
\nonumber 
H & = & \frac{n}{4} \int dx
\left\{ \left[ 1 + (\delta/\pi)^2 \right] \left[(\partial_x \phi_+)^2 +
(\partial_x \phi_-)^2 \right] + \pi^2 \left[\Pi_+^2 + \Pi_-^2 \right]
\right. \\
&&\left.
+ 2\delta \left[ \Pi_- \partial_x \phi_+ - \Pi_+ \partial_x \phi_-
\right]
\right\} \;\;,
\label{eq:heff}
\end{eqnarray}
where $n = 4n_0/L$ is the particle density. Introducing new variables
\begin{equation} 
\tilde{\phi}_\pm = \phi_\pm \quad, 
\quad \tilde{\Pi}_\pm = \Pi_\pm \mp \frac{\delta}{\pi^2} \partial_x
\phi_\mp
\label{eq:nv}
\end{equation}
the Hamiltonian takes an apparently non--interacting form
(eq.(\ref{eq:heff}) with $\delta=0$).  However, the expression of
single--fermion operators\cite{haldane_bosonisation} is changed and
therefore the asymptotic decay law of the single particle Green function
is
obtained as
\begin{equation} 
G_{R\pm}(x) = \langle \psi^{\phantom{\dagger}}_{R\pm}(x) 
\psi_{R\pm}^\dagger(0) \rangle
\approx e^{ i (k_F\pm N_{\mp 0} \delta) x}x^{-1-\alpha} \;\;,
\label{eq:g}
\end{equation}
with $\alpha = \delta^2/(2\pi^2)$. Thus, for any non--vanishing $\delta$
the decay is faster than $1/x$, leading, amongst other things, to the
well--known power--law singularity of the momentum distribution function
at
$k_F$. The correctness of eq.(\ref{eq:g}) can be checked independently
using
the eigenfunctions of eq.(\ref{eq:heig}): one obtains a Toeplitz
determinant
of the form previously considered by Luttinger \cite{luttinger_model},
and
which has the same asymptotic power law as obtained by the bosonization
approach.

Similarly, correlations of two--particle operators decay as $x^{-\eta}$,
with
interaction dependent exponent $\eta$. Specifically:
\begin{eqnarray}
\psi^\dagger_{R\pm} \psi^{\phantom{\dagger}}_{L\pm}  & \Rightarrow &
\eta_1 = 2
\label{eq:exp1} \\
\psi^\dagger_{R\pm} \psi^{\phantom{\dagger}}_{L\mp}  & \Rightarrow & 
\eta_2 = 1+(1 \mp \delta/\pi)^2 \\
\psi_{R\pm} \psi_{L\pm}  & \Rightarrow & \eta_3 = 2 +2 (\delta/\pi)^2 
\label{eq:exp3} \\
\psi_{R\pm} \psi_{L\mp}  & \Rightarrow & \eta_4 = 1+(1 \pm \delta/\pi)^2 
\label{eq:exp}
\end{eqnarray}
The most slowly decaying correlations identify the dominant incipient
instabilities. In the spin language, for positive $\delta$ then {\em
spiral}
spin--density wave correlations and opposite--spin Cooper pairing
correlations with one fixed spin orientation ($\uparrow\downarrow$ and
$\downarrow\uparrow$ are {\em not} degenerate) are favored, whereas for
negative $\delta$ correlations with reversed spin orientations
dominate. Adding a density--density interaction between the two spin
orientations, the degeneracy between pairing and spin--density wave
correlations is lifted. The density correlations, eq.(\ref{eq:exp1}),
are
not affected by the interactions because they are diagonal element of
the
density matrix which themselves are unchanged by the unitary
transformation,
eq.(\ref{eq:uni}). We notice that the exponent for pairing correlations
with
equal pseudospin, eq.(\ref{eq:exp3}), is just twice the exponent of the
single--particle Green function, i.e. there are no singular vertex
corrections in this particular two--particle correlation function.
Finally,
from eq.(\ref{eq:q}) it is clear that the value of $\delta$ is relevant
only
modulo $2\pi$. Consequently, the results (\ref{eq:g}) to (\ref{eq:exp})
are
valid only for $|\delta|\le \pi$. Outside this interval $\delta$ has to
be
taken modulo $2\pi$. We note that {\em the scaling relations between the
different exponents in eqs.(\ref{eq:g}) to (\ref{eq:exp}) are different
from
those of standard fermionic Luttinger liquids} because of the presence
of
time--reversal breaking terms in the Hamiltonian.

We can now comment on Luttinger's original solution of his
model.\cite{luttinger_model} In first--quantized form his Hamiltonian
only
contains first derivatives:
\begin{equation} 
H_{\text{Lutt}} = \sum_{\sigma i} \sigma \Pi_{\sigma i} = \sum_i  p_{+i} 
-\sum_j  p_{-j} + 2\sum_{ij} V(x_{-j}-x_{+i}) \;\;.
\label{eq:hlut}
\end{equation}
We first remark that this Hamiltonian is a conserved quantity as far as
the
Hamiltonian (\ref{eq:h}) is conserved, and shares all (non degenerate)
eigenfunctions. It is unbounded from below though, unlike $H$ in
Eq(\ref{eq:h}). Hence the issue of finding its {\em groundstate} is
replete
with difficulties familiar from relativistic field theories.  The
second--quantized version of the model $H_{\text{Lutt}}$ can be solved
consistently and exactly by filling the Dirac sea and using
bosonization.\cite{mattis_lieb_bos} This leads, amongst other things, to
an
asymptotic decay exponent of the single particle Green function
$\alpha=1/\sqrt{1-(\delta/\pi)^2}-1$. In a first--quantized framework, a
consistent but different solution can be obtained if one is willing to
consider quasi--groundstates where single--particle states below a
certain
very negative cutoff energy $E_{\text{cutoff}}$ are left empty
(evidently,
the model does not have a conventional groundstate). This ``rapidity
cutoff'' in fact is frequently used in the Bethe ansatz solution of
field
theoretical models\cite{bergknoff_mtm_exact,andrei_kondo_review}.  Such
a
state becomes natural if one is interested in finding the {\em
groundstate}
of $H$ in eq.(\ref{eq:h}), and examines the eigenvalue of
$H_{\text{Lutt}}$,
a commuting operator, in this state.  The transformations used for
eq.(\ref{eq:h}) also can be used here, and the solution found as earlier
and
lead to a shift in its eigenvalue due to interactions $\delta
E_{\text{Lutt}}=\frac{2 \pi}{L} \{ N_+ [[\frac{N_- \delta}{ 2
\pi}]]_{rem} +
N_- [[\frac{N_+ \delta}{ 2 \pi}]]_{rem} \} $, a number of the $O(1)$ as
one
would expect from a current carrying state.  The correlation function
can be
found using Luttinger's original paper and lead to the same asymptotic
decay
exponent $\alpha$ of the Green function as in eq.(\ref{eq:g}). This
result
was in fact obtained in Luttinger's paper, \cite{luttinger_model} i.e.
{\em
Luttinger's result in fact applies to the first--quantized solution of
the
model described here}.\cite{rem} The same result for correlation
exponents
can also be obtained by considering variations of the energy with
particle
number, similar to what we described above. We note that the
Mattis--Lieb
and Luttinger results for $\alpha$, though different in general, agree
to
the lowest nontrivial order in $\delta$. The differences at higher order
clearly have to be attributed to the different cutoff procedures used in
the
two calculations.

We now turn to similar models defined on a one--dimensional
lattice. Specifically, we will consider the Hamiltonian
\begin{equation}
H= -  \sum_\sigma \sum_{m=1}^{L}  \;\left[ \exp\left( i \sigma
 \sum_l \alpha_{m-l}  n_{l ,-\sigma} \right)
  c^\dagger_{m\sigma} c^{\phantom{\dagger}}_{m+1\sigma} + h.c. \right]
+ V
\end{equation}
where $\alpha$ is a periodic function, $\alpha_{m+L} = \alpha_{m}$, and
the
number operator $n_m= c^\dagger_m c^{\phantom{\dagger}}_m$.  This
corresponds to a lattice where the up electrons feel a ``gauge
potential''
due to the down electrons, and vice versa.  Here periodic boundary
conditions are implied, i.e. $c_{L+1,\sigma} = c_{1\sigma}$. We notice
that
for the particular case where only $\alpha_0$ is nonzero, this lattice
model
only involves two--particle interactions, contrary to our continuum
model
where some three--body interaction is unavoidable.

The interaction term $ V $ can be either zero, or one of two non trivial
functions which retain the exact solvability. We will consider the XXZ
model
and the Hubbard model, i.e.
\begin{equation}
V_{\text{XXZ}}= V \sum_{j, \sigma} n_{j \sigma} n_{j+1, \sigma}
 \;\;\mbox{and}\;\; V_{\text{Hub}}= U \sum_j n_{j \uparrow} n_{j
\downarrow}
\end{equation}
and will show below that these are exactly  solvable. The XXZ model 
corresponds to two copies of the usual model, wherein the two  
species of particles (spin up and down) only talk to each other
via the phase factors. The  Hubbard  model corresponds to the usual
two body interaction.

We now perform a unitary transformation induced by $U= \exp(i S)$ where
$S=
\sum_{1 \leq l,m \leq L} \beta_{l,m} n_{l \uparrow} n_{m \downarrow}$.
It is
easy to see that $\beta_{i,j} = - \beta_{j,i}$ , i.e. an odd function is
appropriate, and we will assume it to be so.  Thus we find
\begin{equation}
c_{m  \sigma} \rightarrow e^{iS} c_{m  \sigma}  e^{-iS}
= c_{m  \sigma} 
\exp \left[ - i \sigma \sum_l \beta_{m,l}
n_{l  ,-\sigma} \right]
\end{equation}
The transformed Hamiltonian takes the form
\begin{eqnarray}
H' & = & - \sum_\sigma \left[ \sum_{m=1}^{L-1} \exp\left( i \sigma
\sum_l\{
 \beta_{m,l}- \beta_{m+1,l} + \alpha_{m-l} \} n_{l ,-\sigma} \right)
 c^\dagger_{m\sigma} c^{\phantom{\dagger}}_{m+1\sigma} \nonumber \right.
\\
& & - \left. \exp\left( i \sigma \sum_l\{ \beta_{L,l}- \beta_{1,l} +
\alpha_{L-l} \} n_{l ,-\sigma} \right) c^\dagger_{L \sigma}
c^{\phantom{\dagger}}_{1\sigma} + \mbox{h.c.} \right] + V
\end{eqnarray}

We now use the freedom in defining $\beta$ to cancel the interior 
terms in the phase factor by choosing
\begin{equation}
\label{eq:deriv}
\beta_{m+1,l}- \beta_{m,l}= \alpha_{m-l} \;\;  \mbox{for} \; 1 \leq m
\leq
L-1 \; \mbox{and} \; 1\leq l \leq L. 
\end{equation}
The hop across the  $L \leftrightarrow 1 $ bond has a total phase
\begin{equation}
\chi_\sigma=  \sigma \sum_l [ \beta_{L,l}- \beta_{1,l} + \alpha_{L-l} ]
n_{l,-\sigma}.
\end{equation}
It is in fact not necessary to solve explicitly for $\beta$, although it
is
easy enough to do so for simple choices of $\alpha$. We can add eqs.
(\ref{eq:deriv}) for $1 \leq m \leq L-1$ and further add to it
$\alpha_{L-l}$
to find
\begin{equation}
 \beta_{L,l}- \beta_{1,l} + \alpha_{L-l} = \sum_{n=1}^L \alpha_{n-l}
\equiv
 \delta
\end{equation}
since the sum is independent of $l$. This gives $\chi_\sigma =
\hat{N}_{-\sigma} \sigma \delta$. The number operator $\hat{N}_\sigma
\rightarrow N_\sigma$ in any sector, and hence we see that the problem
collapses to one with lattice fermions having twisted boundary
conditions. If $V=0$, we can follow the logic used for the continuum
model
to determine asymptotics of correlation functions. It turns out that up
to
the trivial replacement $v_F = \pi N/2L \rightarrow 2 \sin(\pi N/2L)$
one
obtains the same expression for $E^{(2)}$ as in the continuum limit, and
consequently the same low--energy effective Hamiltonian,
eq.(\ref{eq:heff}),
and the same expressions for correlation exponents (eqs.(\ref{eq:g}) to
(\ref{eq:exp})) apply.

In the presence of a nonzero extra interaction $V$, previous work
\cite{shastry90} can be used where the Bethe Ansatz has been adapted to
the
case of a ``spin twist'', which is precisely the case needed here.  We
write
the solution immediately: for the XXZ model
\begin{equation}
E= -  \sum_{n, \sigma} \cos k_n^\sigma
\end{equation}
and
\begin{equation}
L k_n^\sigma = 2 \pi I^\sigma_n + \sigma \delta N_{-\sigma} +
\sum_m  \theta(k^\sigma_n -k^\sigma_m)
\end{equation}
with the usual phase shift and the usual integers $I_n$. For the Hubbard
model
we have a pair of equations. With
 $E = - \sum \cos k_n$ with $1\leq n \leq N$,
and $N= N_\uparrow + N_\downarrow$ we find 
\begin{eqnarray}
L k_n  & = & 2 \pi I_n + N_\downarrow \delta + 2 \sum_j \mbox{tan}^{-1}[
4 (
\Lambda_j - \sin k_n )/U ] \nonumber \\
2\sum_n \mbox{tan}^{-1}[ 4 ( \Lambda_j - \sin k_n )/U ] &= & 2 \pi J_j -
\delta N + 2 \sum_k \mbox{tan}^{-1}[ 2 ( \Lambda_j - \Lambda_k )/U ].
\end{eqnarray}
In these models, one has non Fermi liquid behavior even in the absence
of
$\alpha_{r}$, and adding this changes the exponents, and indeed even the
symmetries of the model, e.g. for the Hubbard model we have less than
$SU(2)$ invariance. The detailed behavior of the lattice models and the
resulting exponents will be reported elsewhere.

There is a number of further possible generalizations of the present
model:
first, the unitary transformation (\ref{eq:uni}) will in fact put
Hamiltonians containing arbitrary powers of the covariant momenta
$\Pi_{\sigma i}$ into diagonal form. Provided the highest nonvanishing
power
is even, these models have a well--defined groundstate. One could thus
study
models with complicated bandstructures, involving e.g. more than two
Fermi
points. Similarly, in the lattice model certain forms of hopping terms
beyond nearest neighbors can be included.

Another generalization is obtained by giving additional internal degrees
of
freedom to the $\sigma=+$ and $\sigma=-$ particles. For example,
assuming
that they both occur in $m$ different ``flavors'' one obtains a model
with
an internal $SU(m)\times SU(m)$ symmetry. By a calculation analogous to
that
leading to the exponents in eqs.(\ref{eq:exp1}) to (\ref{eq:exp}) one
finds
$\eta_1 =2, \eta_2 = 2 \mp 2\delta/\pi + m \delta^2/\pi^2, \eta_3 = 2 +
2 m
\delta^2/\pi^2, \eta_4 = 2 \pm 2\delta/\pi + m \delta^2/\pi^2$. As
expected
from symmetry, these exponents are independent of the flavor indices
appearing in the corresponding operators. One can further solve the case
where the number of flavors for the $+$ and $-$ particles is different.

In conclusion we have presented a class of lattice and continuum fermion
models which are exactly solvable by a pseudo--unitary transformation,
leading to nontrivial and non--Fermi--liquid behavior, with exponents
depending upon the interaction. The models, unlike those solvable by
bosonization, do not have an unbounded spectrum, and eliminate the
problem
of the negative energy Dirac sea and consequent Schwinger terms, and
help us
to focus on the physics of the interactions in one dimension in a
bounded,
and even a finite dimensional Hilbert space (for the lattice models).
The
method used embeds the original problem considered by Luttinger in a
family
of commuting Hamiltonians which contain both bounded as well as
unbounded
operators. By focusing on the problem of finding the {\em groundstate}
of
the bounded operators one comes up with eigenfunctions which are of the
type
considered by Luttinger, enabling us to make a connection between the
methods used by him (Toeplitz determinants and the Szeg\"o formula for
asymptotics) with more recent conformal/Luttinger liquid methods. The
relatively simple form of the exact wavefunctions may also make it
possible
to understand in detail physical properties in the non--asymptotic
(intermediate and high energy) regime where many questions still remain
open, even in otherwise well--understood one--dimensional models.
  
One of us (H.J.S.) wishes to acknowledge the warm hospitality of the
Indian
Institute of Science, Bangalore, where some of this work was done. This
research was supported in part by the National Science Foundation under
Grant No. PHY94-07194.


\begin{thebibliography}{10}

\bibitem{bethe_xxx}
H.~A. Bethe, Z. Phys. {\bf 71},  205  (1931).

\bibitem{luttinger_model}
J.~M. Luttinger, J. Math. Phys. {\bf 4},  1154  (1963).

\bibitem{mattis_lieb_bos}
D.~C. Mattis and E.~H. Lieb, J. Math. Phys. {\bf 6},  304  (1965).

\bibitem{lieb_hubbard_exact}
E.~H. Lieb and F.~Y. Wu, Phys. Rev. Lett. {\bf 20},  1445  (1968).

\bibitem{tomonaga}
S.~T. Tomonaga, Prog. Theor. Phys. {\bf 5}, 544 (1950).

\bibitem{haldane_bosonisation}
F.~D.~M. Haldane, J. Phys. C {\bf 14},  2585  (1981);
F.~D.~M. Haldane, Phys. Rev. Lett. {\bf 45},  1358  (1980).

\bibitem{bergknoff_mtm_exact}
H. Bergknoff and H.~B. Thacker, Phys. Rev. D {\bf 19}, 3666 (1979).

\bibitem{andrei_kondo_review}
N. Andrei, K. Furuya, and J.~H. Lowenstein, Rev. Mod. Phys. {\bf 55},
331
  (1983); A. Tsvelick and P. B. Wiegmann, Adv. Phys. { \bf 32}, 453
(1983).

\bibitem{rem} Luttinger \cite{luttinger_model} introduced the extra
condition $\bar{V}\equiv \delta =0$. That this condition is unnecessary
for
a formal solution was already remarked by Mattis and
Lieb.\cite{mattis_lieb_bos} Their conclusion that relaxing this
condition
leads to an ill--defined thermodynamic limit for the field theoretic
problem is not very obvious.

\bibitem{shastry90}
B.~S. Shastry and B. Sutherland, Phys. Rev. Lett. {\bf 65},  243
(1990).

\end{thebibliography}
\end{document}